# A new True Triaxial Apparatus with pore fluid system for rock deformation under representative crustal stress conditions

Ashley Stanton-Yonge[1,2], Thomas M. Mitchell[2], Philip G. Meredith[2], Neil Hughes[2], Steve Boon[2], John Browning[3], David Healy[4], and John Bowles[2]

[1]Department of Geosciences, University of Padua, Italy

[2]Department of Earth Sciences, University College London, UK

[3]Department of Structural and Geotechnical Engineering, Pontificia Universidad Católica de Chile, Santiago, Chile

[4]University of Leeds, Leeds, UK

# 1 Introduction

The general state of stress in the crust is triaxial (Zoback et al., 2003). In consequence, rocks experience deformation and eventual failure under the influence of three principal stresses acting along mutually orthogonal orientations, denoted as $\sigma_1$, $\sigma_2$, and $\sigma_3$, with $\sigma_1 > \sigma_2 > \sigma_3$. Despite this, rock deformation experiments are typically conducted under axisymmetric (or conventional triaxial) loading configurations, consisting of cylindrical rock specimens subjected to confining pressure plus axial stress. Due to its versatility, the conventional triaxial test constitutes one of the most widespread experimental procedures in rock deformation studies. However, this configuration disregards the effect of $\sigma_2$ on the physical and deformation properties of rocks, thus complicating the extrapolation of results to crustal conditions.

Based on experiments initially conducted by Kármán (1911) and Boker (1915), and aiming to examine the influence of $\sigma_2$ on rock strength, Murrell (1965) and Mogi (1967) independently conducted conventional triaxial tests employing the two loading configurations permitted by the method: the compressive test, where axial stress exceeds confining pressure ($\sigma_1 > \sigma_2 = \sigma_3$), and the extension test, where confining pressure surpasses axial stress ($\sigma_1 = \sigma_2 > \sigma_3$). A consistent discrepancy between the failure envelops of several types of rocks under the two loading schemes indicated that the magnitude of $\sigma_2$ with respect to $\sigma_1$ significantly affects rock strength. Similar results were obtained by Handin et al. (1967), who performed pioneering experiments of rocks subjected to truly triaxial states of stress by employing a combination of compression and torsion on rock specimens of hollow cylindrical shape.

Since then, multiple efforts have been conducted to design and implement rock deformation apparatuses to allow for independent control of three loading axes (Takahashi & Koide, 1989; Mogi, 2006; Li et al., 2012). Takahashi et al. (2001) established three categories of True Triaxial Apparatus (TTA) depending on the loading method. The rigid platen TTA (Furuzumi & Sugimoto, 1986; Esaki & Kimura, 1989; King et al., 1997; Skoczylas & Henry, 1995; Sibai et al., 1997; Alexeev et al., 2004; Tiwari & Rao, 2004; Cheon et al., 2006; He et al., 2010; Tshibangu & Descamps, 2012; Rukhaiyar & Samadhiya, 2017) consists of either three or six loading rams transmitting stresses to a cubic or prismatic rock specimen through rigid platens. This configuration allows for high loading capacities and testing specimens with large dimensions. However, the installation of three or six loading rams with an appropriately stiff reaction ring and loading frame makes this design considerably complex and expensive. In addition, the rigid platens configuration induces loading boundary effects that may cause significantly heterogeneous stress distributions within the sample. First, the stiffness contrast between the rock sample and the rigid platens generates radial shear forces at the rock-platen interface, causing a clamping effect within a region at the edges of the sample (Mogi, 1967; Hawkes

& Mellor, 1970). Second, the requirement that loading platens must necessarily be slightly smaller than the rock specimen results in unstressed sample edges that may cause significant stress concentrations (Shi et al., 2012).

By contrast, flexible medium apparatus (Atkinson & Ko, 1973; Michelis, 1988; Smart, 1995) achieve genuinely triaxial stress states on rock samples by employing fluid pressure along two or three loading axes. Loads are transmitted uniformly to the surfaces of the rock sample through flexible cushions. This design allows for homogeneous stress distributions in the sample, effectively avoiding unstressed edges and friction end effects. However, the loading capacity of such designs is usually not sufficient for studying rock deformation and failure.

A third type of TTA, initially developed by Mogi (1971) and thereafter adapted in several laboratories (Spetzler et al., 1981; Takahashi & Koide, 1989; Wawersik et al., 1997; Haimson & Chang, 2000; Lombos et al., 2012; Feng et al., 2016; Shi et al., 2017), employs the rigid platens method to load the sample in the $\sigma_1$ and $\sigma_2$ directions and a fluid confining medium for $\sigma_3$. This configuration can achieve high loads and, by applying $\sigma_3$ homogeneously, loading boundary effects are reduced (Mogi, 2006). In addition, because the design already contemplates subjecting the sample to confining pressure, the introduction of a pore fluids system to deform saturated rocks and measure permeability is relatively simple. However, this configuration poses additional challenges, such as the isolation of the rock specimen from the confining medium.

The number of TTA constructed in laboratories have vastly increased during the past decades, as well as the volume of collected data. Although advances towards standardizing the procedures for true triaxial rock deformation tests have been made (Feng et al., 2019), significant issues still remain regarding the repeatability and generalization of results. For example, even though numerous studies have supported the dependence of rock strength on $\sigma_2$, no general consensus on the form of this dependence has been reached (Paterson & Wong, 2005; Kwaśniewski, 2012; Xie et al., 2022). Furthermore, with some exceptions (Lombos et al., 2012; Shi et al., 2017), the vast majority of the implemented TTA do not consider a pore fluids system and can only deform dry rock. As a result, there is currently an important gap in the knowledge of deformation of saturated rocks and their fluid flow properties under representative crustal stress conditions.

Aiming to study deformation and failure of rocks under true triaxial conditions, the design and manufacture of a new apparatus was commissioned in the Rock and Ice Physics laboratory of UCL in 2002. In its initial design, the apparatus consisted of a pressure vessel and 4 double acting actuators ('X' and 'Y' axes) located in the vessel wall, to load a square section rock specimen within the pressure vessel, plus two additional actuators at the bottom and top of the vessel to provide axial load along the Z axis. The original conceptualization of this apparatus included two different operating modes: (1) using 30x30x75mm samples, and (2) using 50x50x50mm cubic samples. The prismatic samples arrangement was envisioned to examine true triaxial deformation and failure using the normal 2.5 to 1 length to width ratio needed for strength and failure measurements. The cubic sample arrangement was designed to investigate the evolution of the crack damage envelope under true triaxial loading conditions, as reported by Browning et al. (2017) and Browning et al. (2018), but under saturated and elevated pore fluid pressure conditions, that is, under conditions more relevant to those at depth in the Earth's crust.

A major challenge of deformation experiments of cubic rocks loaded through six pistons and under pressurized pore fluid conditions is the sealing of the samples' exposed edges. A jacket system is needed to isolate the pressurized pore fluids within the sample from the confining medium, a challenging feat due to the sharp edges of the cubic sample. The initial concept of UCL's TTA was to address this issue by dispensing with edge seals altogether, and to saturate and pressurize the sample with the fluid inside the pressure vessel (water or brine). Since this initial concept, the system evolved through the introduction of an edge sealing system and two fluid pressure pumps that allow to measure fluid flow across the sample whilst being deformed under true triaxial conditions.

In its current form, the TTA instrument manufactured in UCL consists of six servo-controlled loading rams that transmit maximum stresses of 220 MPa in the two horizontal axes and 400 MPa in the vertical axis to 50 mm side cubic rock samples through rigid platens. The sample and loading platens are introduced in a steel vessel where rock specimens can be subjected to up to 60 MPa of confining pressure driven by silicone oil. Pore fluids lines are installed along the three loading axes, which are connected to two pump intensifiers that allow for highly accurate permeability measurements by means of the steady state method.

The present article is organised as follows. We first describe the implementation and design of UCL's TTA, followed by details on load calibrations. Next, we present a suite of Finite Element Method (FEM) models implemented to determine the conditions and loading configuration that minimise the loading boundary effects during true triaxial loading. These observations are generic and we therefore expect they will contribute to the development of true triaxial loading systems generally. Finally, we validate our experimental configuration by presenting results on permeability measurements along the three axes on cubic samples of three types of well studied rocks: Darley Dale Sandstone (DDS), Crab Orchard Sandstone (COS), and Etna Basalt (EB).

## 2    Design and implementation of True Triaxial Apparatus

### 2.1    Loading frame and control system

A high stiffness loading frame made of steel alloy was designed in-house to withstand the forces acting along the vertical (Z) axis to up to 1500 kN. A reaction ring was designed and constructed to provide support for the forces acting along the two horizontal axes (X and Y). One servo valve is used for controlling the two hydraulic rams of each loading axis (Figure 3). The load applied along the horizontal axes is measured through differential pressure transducers which record the difference between the hydraulic ram's power and return lines. The load along the vertical axis is measured by means of an external load cell installed on the upper piston.

### 2.2    Confining and pore pressure system

A steel vessel was designed in-house to withstand confining pressure of up to 60 MPa. The confining fluid is silicone oil, and the pressure is applied through one double-acting intensifier. The pore fluid system (Figure 4) comprises two ISCO pumps that accurately control the pore fluid (distilled water) pressure. Permeability is measured by applying a pressure difference between the two pumps and allowing steady-state flow to develop along one of the three loading axes at a time. Two independent flow measurements are accurately taken by each pump, and permeability is calculated by applying Darcy's Law. Once accurate flow measurements are obtained for a certain stress state along one loading axis, flow is interrupted along the recorded direction and is started along the next one. The selection of the axis along which pore fluids are flowing is controlled through six air valves are installed on the pore fluid lines, which can be opened or closed via the virtual interface of the control software (Labview).

### 2.3    Sample assembly and jacketing system

Rock samples are saw cut and grinded into 50 mm side cubes with mutually orthogonal surfaces up to a tolerance of $\mp 0.01$mm by means of a high precision vice and square. The sample assembly was designed to ensure the alignment of the rock specimen throughout the experiments. A cross section view of the constituent parts of the vessel top plug, vessel cap, and sample assembly can be visualised in Figure 5a. The jacket is manually made by applying several layers of liquid Polytek rubber to the exposed edges of the sample, which solidifies after a curing time of approximately 24 hours (Figure 5b). The horizontal loading platens are held in position by an aluminum holder. The bottom piston is kept in the same position throughout the experiments to ensure that the piston caps on the horizontal loading platens are aligned with the cross-bore pistons.

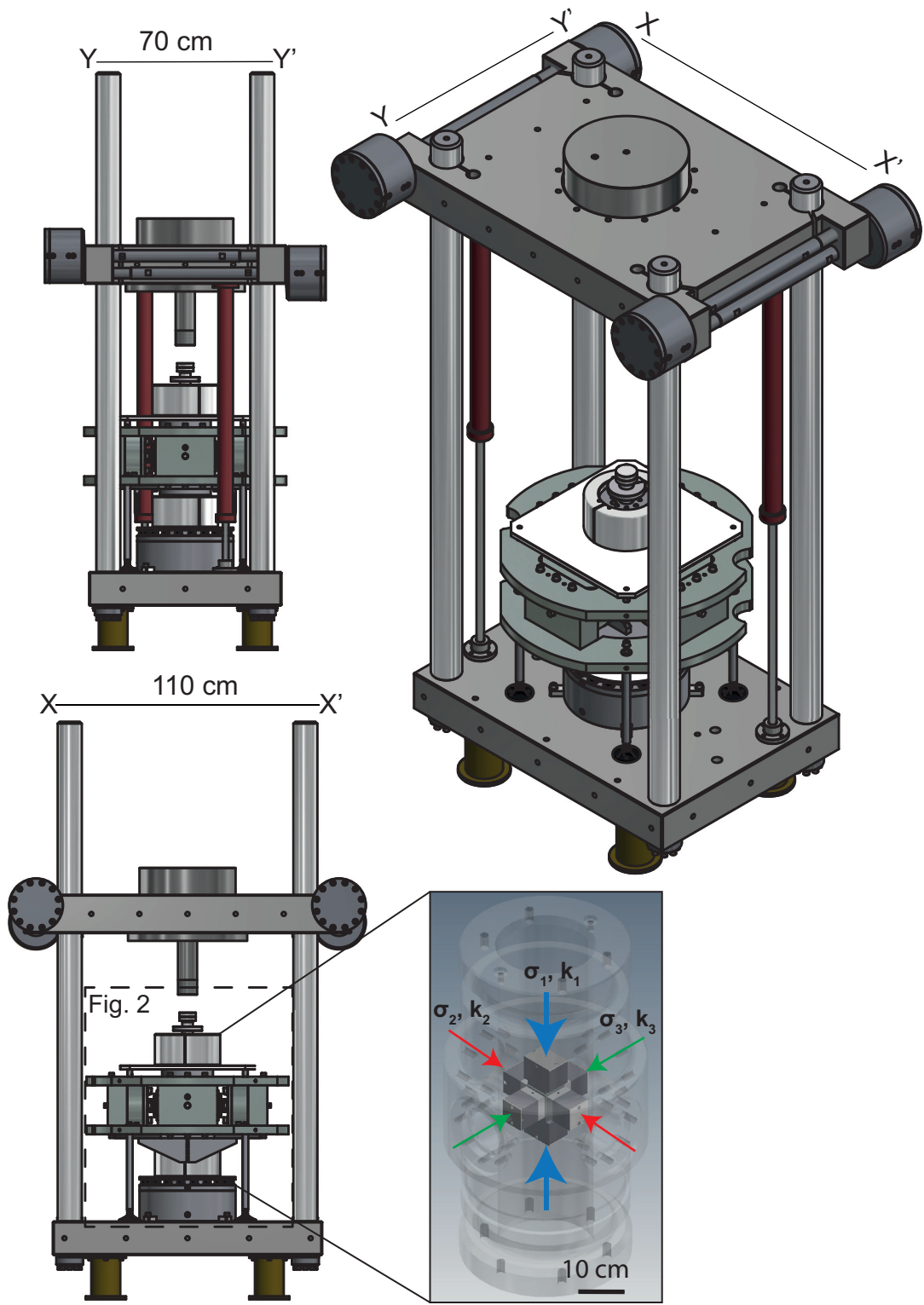

**Figure 1:** Lateral and 3D views of the TTA load frame and reaction ring. The dashed rectangle indicated the section displayed in Figure 2.

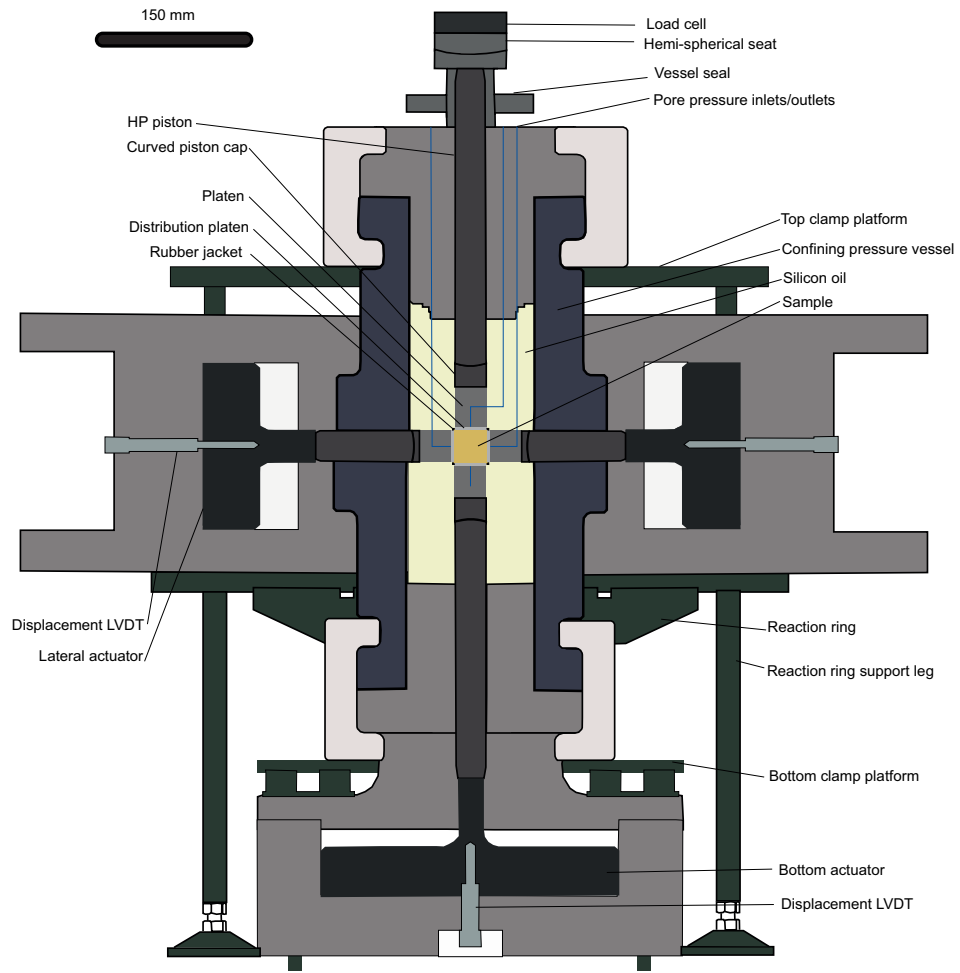

**Figure 2:** Section view of UCL's TTA cell, with labelled constitutive parts.

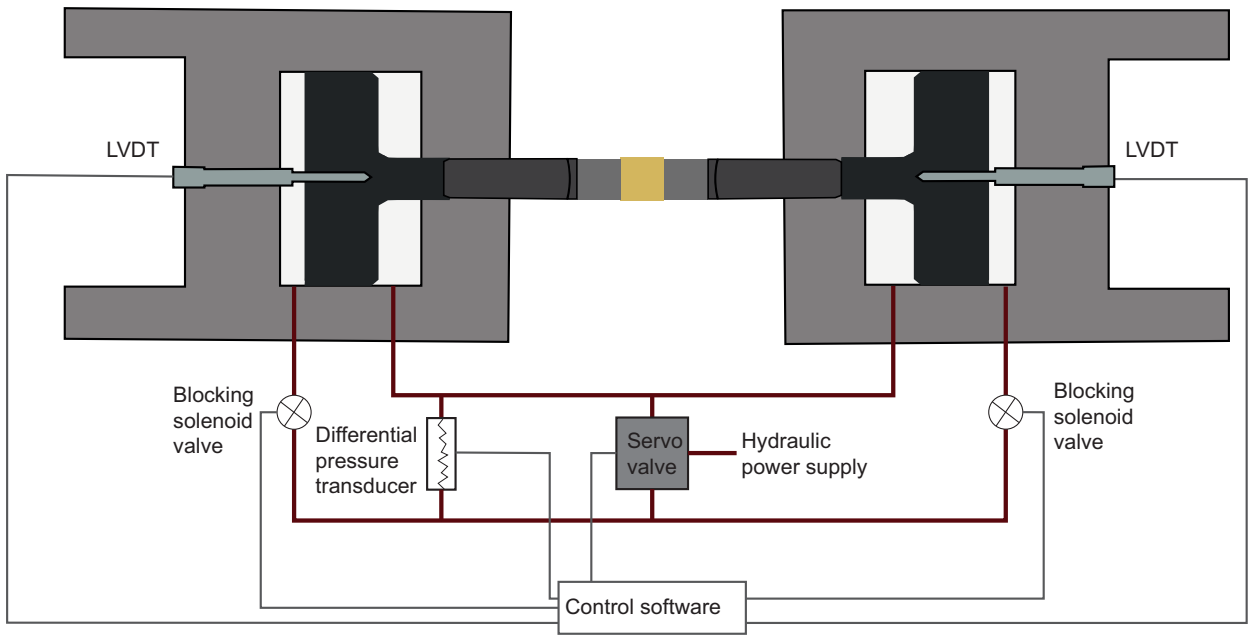

**Figure 3:** Schematic diagram of one loading axis indicating the servo-controlled system.

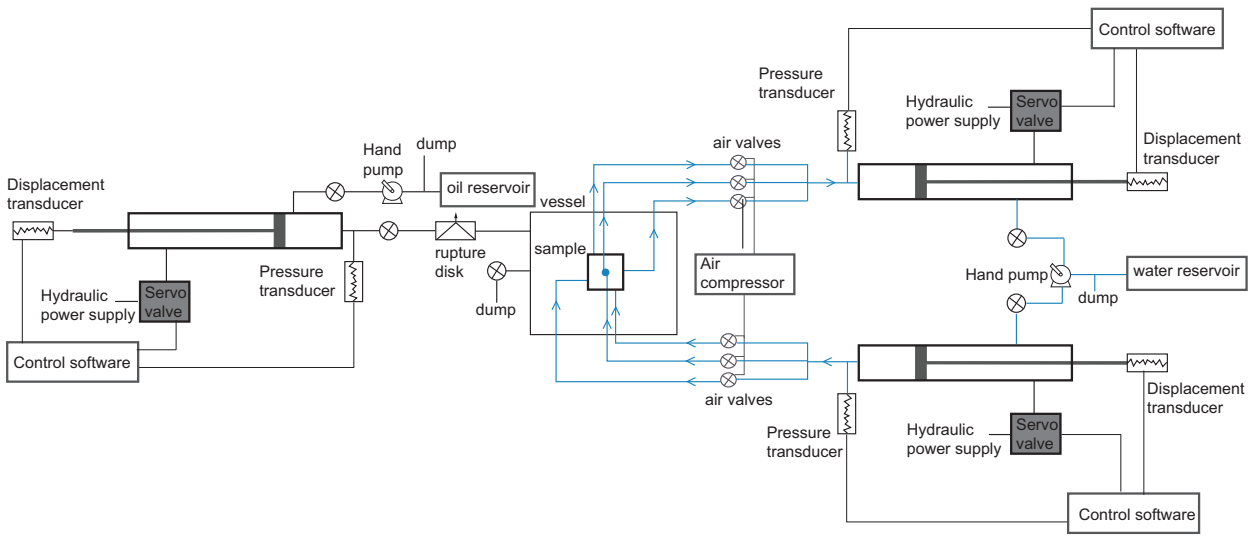

**Figure 4:** Schematic diagram for confining pressure and pore fluids system.

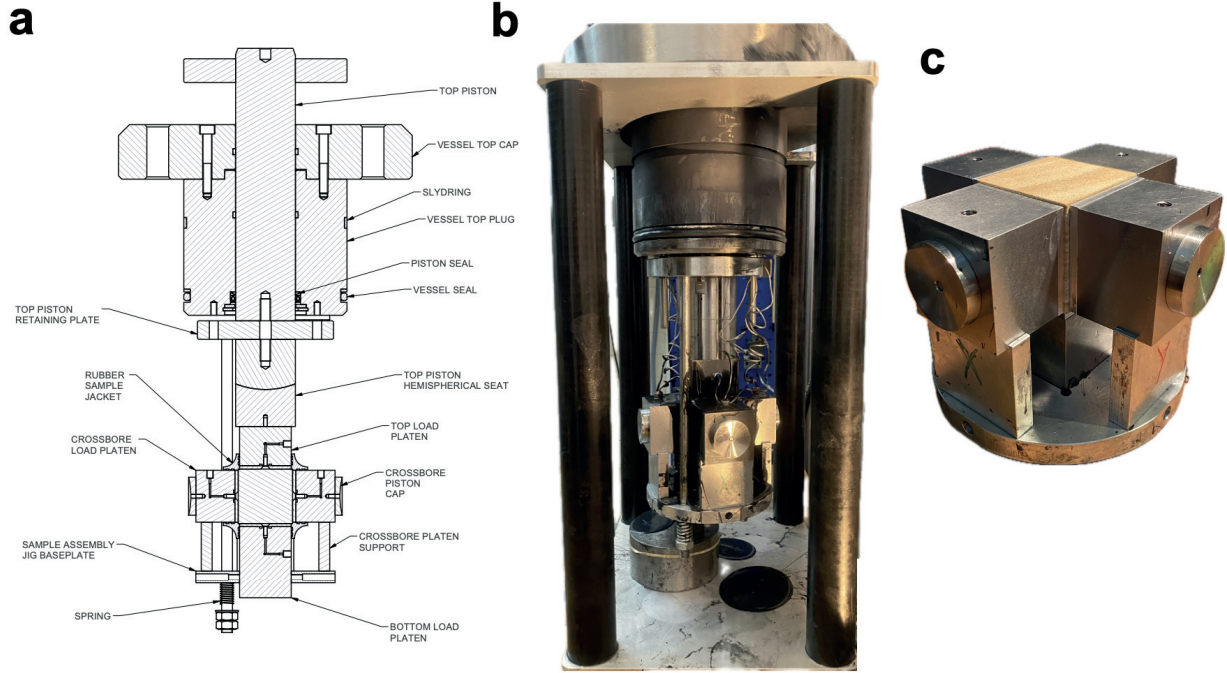

**Figure 5:** (a) Cross-section of the sample assembly and vessel cap design, indicating constitutive parts. (b) Sample assembly with cured rubber jacket in place, and pore fluid lines connected to load platens. (c) A cubic sample of Crab Orchard Sandstone in aluminum holder, prior to the manufacture of the rubber jacket, and without the upper loading platen.

## 2.4 Load calibration

As previously indicated, the apparatus contemplates only one external load cell, attached to the upper piston on the Z axis (Figure 2). Measurements of the load transmitted along the X and Y axes are obtained from differential pressure transducers. Calibrations to the raw measurements obtained by the transducers are determined by applying sequential steps of confining pressure while the vessel is vacant from a sample and sample assembly, but with the pistons introduced inside the vessel. This allowed to compare the readings from the differential pressure transducers to the load that is theoretically being applied to each piston, calculated as $L_{theo} = P_c A_p$, where $P_c$ is the confining pressure and $A_p$ the area of the piston (Figure 6). The external load cell in the Z axis provided an independent and accurate measurement of the applied load. The difference between this reading and $L_{theo}$ allowed us to estimate the seal friction of the piston, which is assumed to be approximately representative of the seal friction of all the pistons in the system.

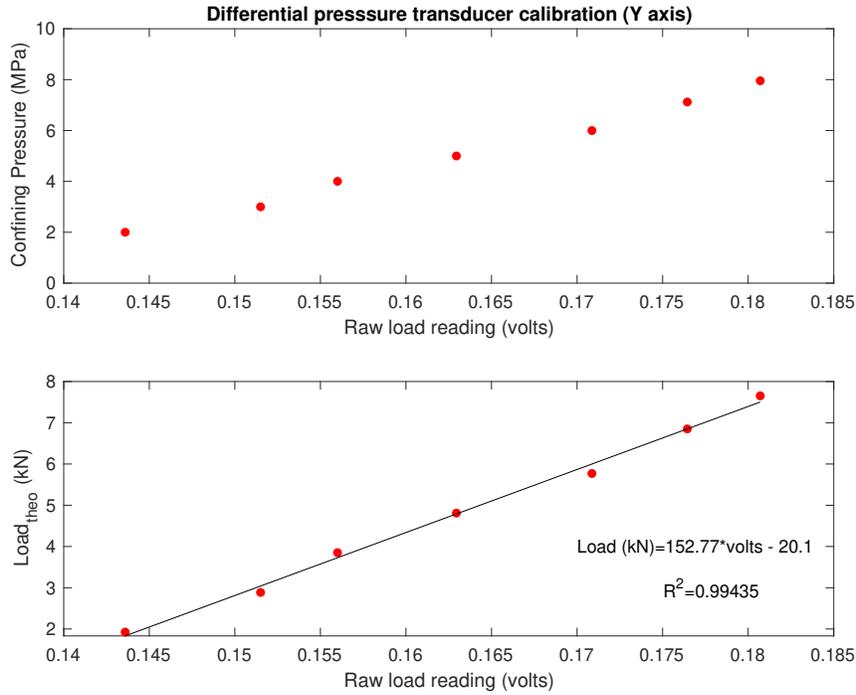

**Figure 6:** Calibration of the differential pressure transducer for load reading along the Y axis.

# 3   FEM models to evaluate loading boundary effects during true triaxial loading

The following section describes the implementation and results of a set of Finite Element Method (FEM) models designed to evaluate the effectiveness of practical measures than can be taken for reducing loading boundary effects in true triaxial rock deformation experiments. In particular, we explore the capability of platen material selection for reducing the end friction effect, and the influence of confining pressure for decreasing the blank corner effect.

## 3.1   Platen material selection for reducing end friction effect

The ideal conditions for rock deformation experiments require that stresses and deformation remain uniform across the loading planes at all times, with no shear stresses or distortion at the boundary planes (Hawkes & Mellor, 1970). However, using rigid pistons and platens (typically made of steel) to transmit stresses to rock specimens creates a stiffness mismatch between the rock and metal. Friction at the rock-platen interface leads to lateral restraint and radial shear forces, causing a clamping effect near the contact areas (Filon, 1902). This results in a heterogeneous stress distribution, which influences the measured mechanical properties of rocks. Stress concentrations can locally weaken the material, while fracture propagation is inhibited in the clamping region (Mogi, 2006).

It has been empirically demonstrated that using a minimum height to diameter ratio of 2.5 for cylindrical samples effectively eliminates the influence of the end friction effect for measuring the uniaxial compressive strength of rocks (Mogi, 1966). For true triaxial tests, though, not only the sample geometry is restricted, but also the shear stresses will develop in each of the four or six rock-platen interfaces, potentially interacting with each other.

Several measures have been proposed to reduce the end friction effect during rock deformation experiments. Decreasing the platen-rock friction coefficient through the use of lubricants and/or teflon sheets is a widely employed method (Shi et al., 2012; Feng et al., 2017). A problem with this method is the intrusion of liquid or soft material into the sample while being compressed, locally decreasing the effective stress and influencing the failure mode and strength of tested materials (Mogi, 2006). The use of a thin copper jacket can prevent this. However, the metal-teflon-copper-rock configuration is impractical for experiments involving fluid flow across the sample due to the requirement of evenly distributed fluid across the rock face to satisfy the assumptions of Darcy's Law.

An alternative method for end friction effect reduction is selecting a material for the platens that matches the elastic properties of the rock specimen. In the configuration of UCL's TTA (Figures 2, 5a) the rock-platen interface is between the sample and a thin platen for fluid distribution across the sample face, attached to the steel loading platen. We therefore use FEM models to explore the effect of different material selection (steel and aluminum) for the manufacture of the fluid distribution platens on the end friction effect developed when deforming a cubic sample of sandstone.

The models were implemented and ran using the Static Structural module of the commercial software ANSYS. The model geometry is based on the drawings and 3D model developed during the design of UCL's TTA. The elastic properties of the materials are indicated in Table 1. The rock sample and platens are modelled as rigid, linear elastic bodies, with a frictional contact between the rock sample and platen described by a static friction coefficient. Because we do not attempt to control or reduce the friction coefficient during our experimental design, the friction coefficient is assumed to be equal for all interfaces and equivalent to 0.2.

| Material | E (GPa) | $\nu$ | Frictional contact coefficient |
|---|---|---|---|
| Sandstone | 35 | 0.2 | 0.2 |
| Steel | 210 | 0.3 | 0.2 |
| Aluminum | 77 | 0.33 | 0.2 |

**Table 1:** Elastic properties of the materials considered for the FEM models.

To validate our model configuration, we first reproduce the end friction effect developed on a cylindrical sample of sandstone under uniaxial compression. This configuration has been analytically evaluated by previous researchers (Balla, 1960), allowing to compare our results with those available in the literature (Figure 7). Results are plotted in the same form as Hawkes & Mellor (1970) (Figure7a), equivalent to the McClintock-Walsh failure criterion ($C$) normalized by the applied stress $\sigma$. $C$ is defined by,

$$C = \frac{\sigma_1}{4}[(\mu^2 + 1)^{1/2} - \mu] - \frac{\sigma_3}{4}[(\mu^2 + 1)^{1/2} + \mu],\tag{1}$$

where $\sigma_1$ and $\sigma_3$ are the maximum and minimum principal stresses developed in the loaded sample, and $\mu$ is the internal friction coefficient of the rock (assumed of 0.7). This failure criteria predicts that cracking will begin when $C$ surpasses the uniaxial tensile strength or the rock. Cracking and the highest crack density will therefore occur in the region of the sample where $C$ is highest.

Our model results (Figure 7b) reproduces the same features from those obtained by Balla (1960) and plotted by Hawkes & Mellor (1970). Mainly, a clamped, low stress region of half elliptical shape in the sample edges, and stress concentrations (shaded/red areas) in the sample corners and in regions in between the clamped shadow and the sample center.

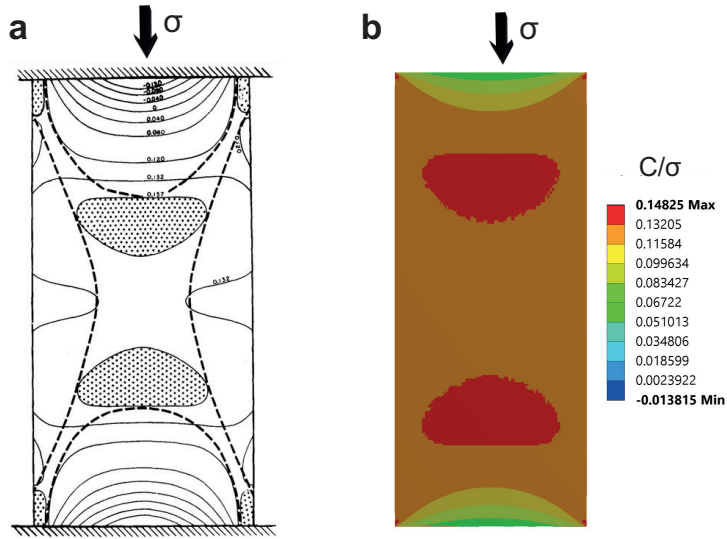

**Figure 7:** (a) End friction effect in uniaxial compression of a cylinder , as plotted by Hawkes & Mellor (1970). Contours indicate the McClintock-Walsh failure criterion ($C$) normalized by the applied stress $\sigma$. Shaded areas indicate highly stressed regions were cracking is likely to initiate. (b) Results by our FEM model, with colormap displaying $C/\sigma$.

We then implement a similar model but with a sandstone sample of cubic shape, loaded through rigid steel platens and fluid distribution platens of variable material (Figure 8a). Results are also presented as colormaps of the $C$ failure criteria normalized by the applied stress $\sigma$ (Figures 8b,c). The sample is loaded only along the vertical axis to visualise the effect of the changing platen material more clearly.

We observe that the use of aluminum platens, with a platen-rock stiffness ratio of $E_{platen}/E_{rock} = 2.1$ reduces the extent of the end friction effect with respect to using steel platens $E_{platen}/E_{rock} = 5.9$ in that the clamped area is reduced, and the stress distribution within the remaining of the sample is considerably more homogeneous.

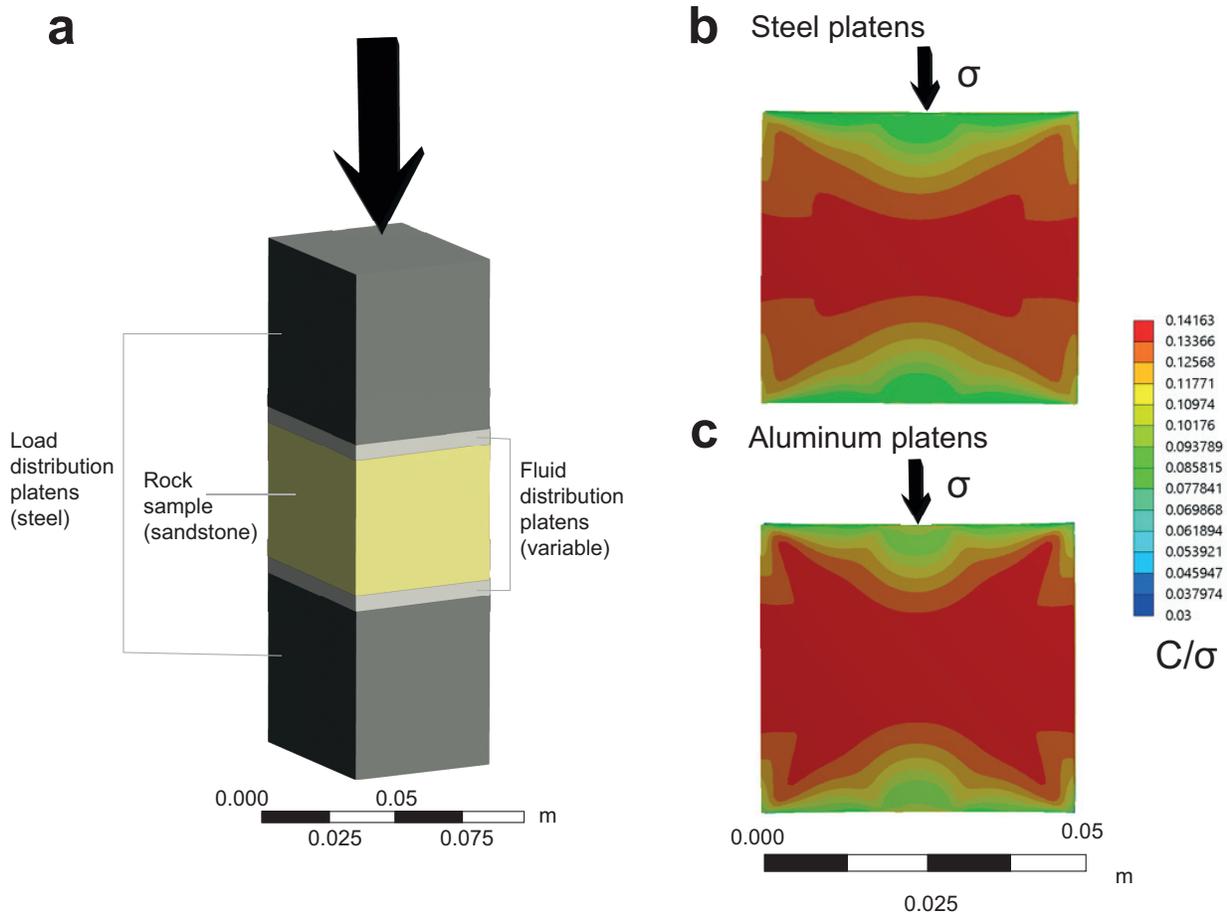

**Figure 8:** (a) Configuration of FEM model to evaluate the effect of platen material selection on the end friction effect of a cubic sample loaded uniaxially. (b) Stress distribution across a sample loaded through steel platens, plotted as the failure criterion $C$ normalized by the applied stress $\sigma$. (c) Stress distribution across a sample loaded through aluminum platens.

## 3.2 Confining pressure for reducing the unstressed edges effect

Another issue inherent to the true triaxial loading of cubic or prismatic rocks arises from the requirement that the loading platens being slightly smaller than the rock sample to allow for the sample to deform while avoiding that the rigid platens interact with each other. This configuration results in that the edges of the samples remain unstressed throughout the experiments, which causes stress concentrations that can severely influence the failure mechanism and strength of tested materials (Shi et al., 2012).

A potential advantage of the Mogi-type TTA with respect to the rigid platens TTA is that the confining medium should help reduce the stress concentrations caused by unstressed edged effect (Mogi, 2006). However, the magnitude of this reduction as a function of the applied confining pressure remains unclear.

We therefore implemented a series of FEM models to quantify the effect that different magnitudes of confining pressure have on the stress concentrations produced by unstressed edges during true triaxial loading. To isolate this effect from other loading boundary effects (i.e. the end friction), the friction coefficient in the platen-rock interface was reduced to 0.1. All the modelled samples are subjected to a maximum stress ($\sigma_1$) of 100 MPa across the x-axis and an intermediate stress ($\sigma_2$) of 50 MPa across the y-axis. The application of $\sigma_3$ varies in the different models. In model 2a (Figure 9) a $\sigma_3$ of 10 MPa is applied through rigid platens across the z-axis, whereas for models 2b and 2c, $\sigma_3$ is applied through a confining medium of 10 and 30 MPa, respectively.

Results are plotted as the $\sigma_2$ across the loaded samples calculated from the models, normalized by the expected $\sigma_2$, along a cross-section of the $\sigma_1/\sigma_3$ plane. Positive compressive stress convention is used. An uniform distribution close to 1 would represent the most ideal conditions for a true triaxial test. We observe that the application of a confining pressure ($P_c$) of 10 MPa (model 2b) significantly reduces the stress concentrations at the edges and corners with respect to the model with $P_c = 0$ (model 2a). The unstressed edges effect is almost completely diminished by the application of $P_c = 30$ MPa (model 2c).

In order to clearly visualize the effect of $P_c$ on the stress concentrations produced by unstressed edges, the magnitude of the most stressed area (negative values representing tensile stresses) of each model is plotted against the modelled $P_c$ (Figure 10). It can be observed that the tensile stressed regions normalized by the expected magnitude of $\sigma_2$ exhibits a linear relationship with the applied confining pressure, and that a $P_c$ equal to 20 MPa eliminates tensile stresses in the unstressed regions. Projection of the linear trend indicates that increasing $P_c$ further would completely eliminate the unstressed edges stress concentrations, by eventually reaching a magnitude of calculated $\sigma_2$ normalized by expected $\sigma_2$ equal to 1.

## 4 Experimental validation

Cubic samples of three highly studied rocks: Darley Dale Sandstone (DDS), Crab Orchard Sandstone (COS), and Etna Basalt (EB), were introduced into UCL's TTA for triaxial permeability measurements under a hydrostatic confining pressure of 5 MPa. Results are consistent with previously reported permeability measurements (Zhu & Wong, 1997; Gehne & Benson, 2017; Vinciguerra et al., 2005). Results from COS are of particular interest, since a significant anisotropy in permeability between the directions parallel and orthogonal to bedding had been previously reported (Gehne & Benson, 2017), which was also detected by our measurements.

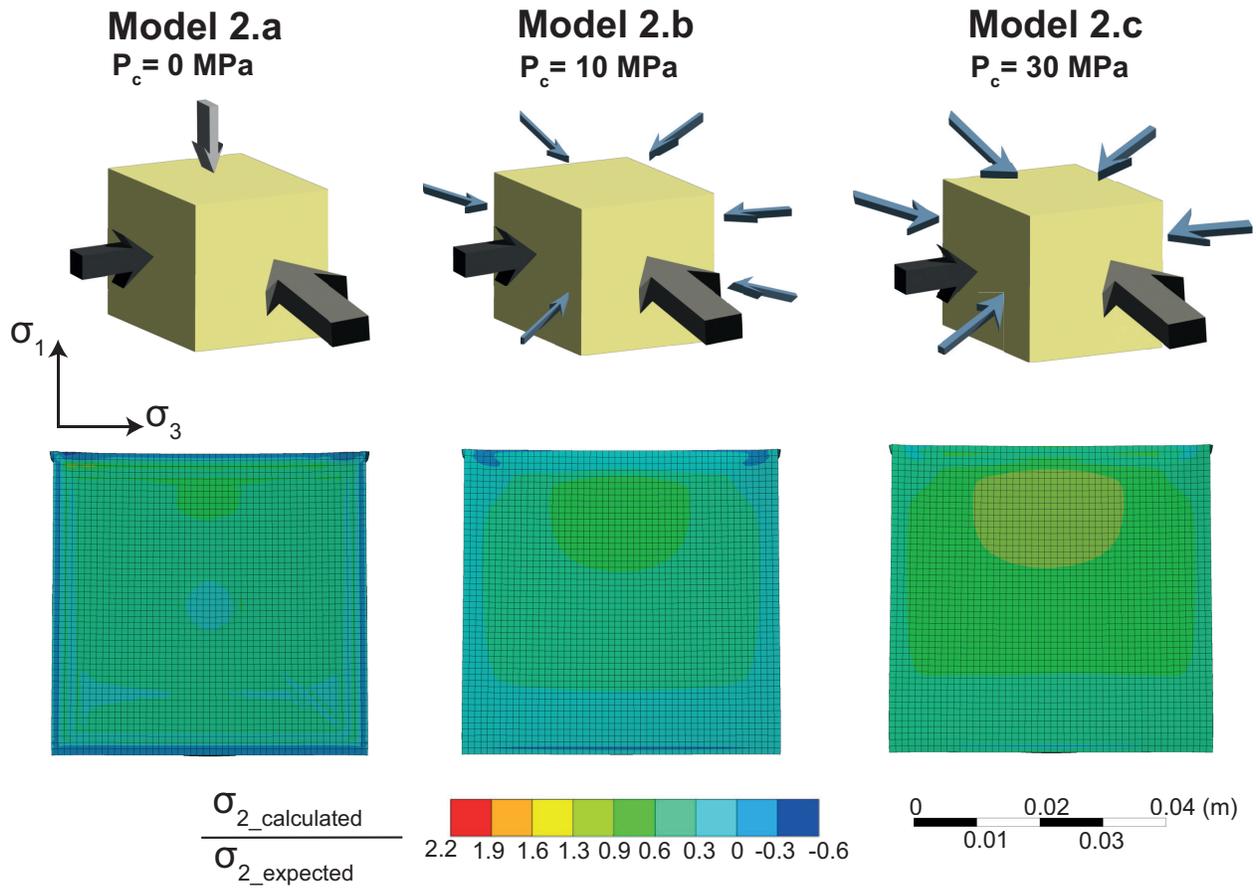

**Figure 9:** Configuration and results of models implemented to quantify the effect of confining pressure on the reduction of stress concentrations due to unstressed edged effect. Positive compressive stress convention is considered.

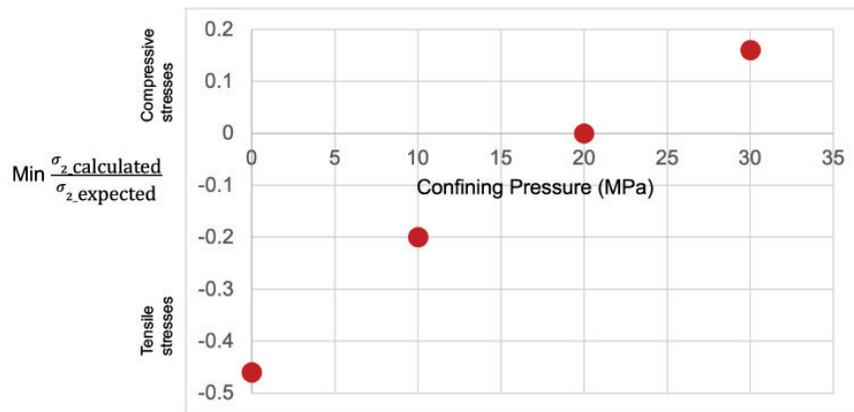

**Figure 10:** Magnitude of the most stressed region of each model (negative magnitudes indicate tensile stresses) as a function of the confining pressure applied in the model.

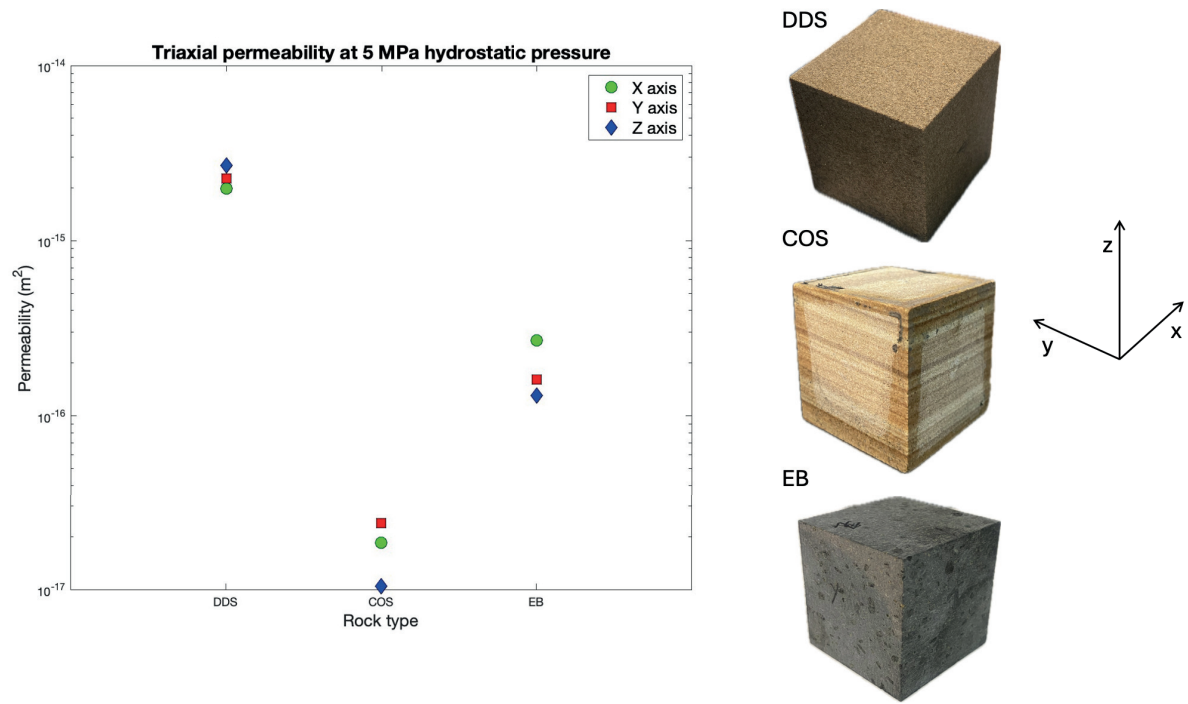

**Figure 11:** Triaxial, hydrostatic (5 MPa) permeability of three well known rocks: Darley Dale Sandstone (DDS), Crab Orchard Sandstone (COS), and Etna Basalt (EB).

# 5 Summary and future perspectives

We have developed a new True Triaxial Apparatus in the UCL Rock and Ice Physics laboratory that successfully transmit independent stresses along three mutually orthogonal axes to 50 mm side cubic samples, whilst accurately measuring the permeability along the three loading axes through the steady-state method. This apparatus will permit producing novel and much needed data on the physical and fluid flow properties of rocks subjected to truly triaxial states of stress, representative of crustal conditions.

We have also explored configurations and experimental protocols that can reduce the loading boundary effects during true triaxial loading through the implementation of a suite of FEM models. We conclude that the selection of the platen material influences the end friction effect. In particular, a material that resembles the elastic properties of the rock sample ($E_{platen}/E_{rock} = 2.1$ for aluminum and sandstone) effectively reduces the extent of the end friction effect with respect to steel ($E_{platen}/E_{rock} = 5.9$), by achieving a considerably more homogeneous stress distribution. In addition, we have quantified the effect that confining pressure exerts on reducing the stress concentrations produced by unstressed edged, finding that a confining pressure of 20 MPa eliminates tensile stresses on the sample corners for the loading conditions considered in our models. We expect that these results will contribute towards developing protocols to standardize the procedures during true triaxial testing, to improve the repeatability of TTA data and expand the development of the method.

## Acknowledgements

This research was supported financially by the NERC grant: "Quantifying the Anisotropy of Poroelasticity in Stressed Rock", NE/N007826/1 and NE/T00780X/1.

## Conflict of interest

The authors have no conflict of interest to declare.